\documentstyle[prl,aps,floats,twocolumn,epsfig]{revtex}

\begin{document}

\twocolumn[\hsize\textwidth\columnwidth
\hsize\csname @twocolumnfalse\endcsname

\title{Strong disorder fixed point in absorbing
state phase transitions}
\author{Jef Hooyberghs$^{1,2}$, Ferenc Igl\'oi$^{3,4}$ and Carlo
Vanderzande$^{1,5}$}
\address{
$^1$ Departement WNI, Limburgs Universitair Centrum, 3590 Diepenbeek, Belgium
\\
$^{2}$ Laboratorium voor Vaste-Stoffysica en Magnetisme, 
Celestijnenlaan 200D, 3001 Heverlee, Belgium \\
$^3$ Research Institute for Solid State Physics and
Optics, PO Box 49,
H--1525 Budapest, Hungary \\
$^{4}$ Institute of Theoretical Physics, Szeged University,
H--6720 Szeged, Hungary\\
$^5$ Instituut voor Theoretische Fysica, Celestijnenlaan 200D, 3001
Heverlee, Belgium}
\date{\today}
\maketitle

   \begin{abstract}
The effect of quenched disorder on non-equilibrium phase transitions in the
directed percolation universality class is studied by a strong disorder
renormalization group approach and by density matrix renormalization group
calculations. We show that for sufficiently strong disorder
the critical behavior is controlled by a strong disorder fixed point
and in one dimension the critical exponents are conjectured to be exact:
$\beta=(3-\sqrt{5})/2$ and $\nu_{\perp}=2$.
For disorder strengths outside the attractive region of this fixed 
point,
disorder dependent critical exponents are detected.
Existing numerical results in two dimensions can be interpreted within a
similar scenario.\\
\ \\
PACS number(s): 05.70.Ln, 05.70.Jk, 64.60.Ak
\end{abstract}

\hspace{.2in}
]

Stochastic many particle systems with a phase transition into an absorbing
state are of wide interest in physics, chemistry and even biology
\cite{MarDic}. Much recent work has focused on establishing a classification
of possible universality classes for systems having this type
of transition \cite{Dic}. For
models with a scalar order parameter,
absence of conservation laws, and short range
interactions, the critical behavior is conjectured to be that of directed
percolation \cite{Grass}. Well known models with a phase
transition in this universality
class are the contact process \cite{Harris} and the
Ziff-Gulari-Barshad model of catalytic reactions \cite{ZGB}.
When there is a conservation law present,
other universality classes can
appear, the best known of which is the parity conserving class 
\cite{CardyTauber}.

In this Letter we study the effect of spatially quenched
disorder
on the directed percolation universality class
and show that for strong enough disorder a new universality class
of absorbing state phase transitions appears.

There are two main reasons for performing such a study. Firstly,
directed percolation has been linked to several experimental situations
such as catalytic reactions \cite{ZGB}, depinning transitions
\cite{Barabasi}, and the flow of granular matter \cite{Sand}
(for a review, see \cite{Hinexp}). So far however,
the directed percolation exponents have not shown up in any of these experimental
realisations. It has been suggested \cite{Hinexp} that the
presence of some form of quenched disorder is responsible for
this discrepancy. Secondly, while several studies have been
performed on the effect of spatially quenched disorder \cite{duitser,BramDur,weerDick,Janssen,Cafiero,Webman},
no coherent picture of these effects has been presented. Here,
we present such a scenario by relating the critical behaviour of the
strongly disordered
contact process to that of a class of random quantum spin chains, such as
the random transverse Ising model. Our results mostly concern
the one-dimensional case, but we will argue that a similar
picture also holds in $d=2$.

We consider the contact process, in which each site of a lattice
can either be vacant ($\emptyset$) or be occupied by one particle
($A$).
The dynamics of the model is a continuous time Markov process in which
particles at site $k$ can disappear ($A \to \emptyset$) with a rate $\mu_k$,
while new particles can be produced on empty sites ($\emptyset
\to A$), with a rate $p\lambda_k/2$ where $p$ is the number of
occupied neighbors. Here both $\mu_{k}$ and $\lambda_{k}$ are
independently and identically distributed random variables
with distributions $\pi_0(\mu)$ and $P_0(\lambda)$, respectively.
Earlier work on random versions of the contact process and
of directed percolation focused mainly on dynamical aspects,
by studying the properties of growth starting from a single seed
particle. At criticality, a breakdown of dynamical scaling was
observed \cite{weerDick}, consistent with indications from a field
theoretical study \cite{Janssen}. The absorbing phase was found to
have properties similar to that of a Griffiths phase and shows
power law behaviour with non-universal exponents \cite{weerDick,Cafiero}.
The static critical behaviour of the model has been explored less,
but exponents in $d=2$ have been determined in
\cite{weerDick}.

In this Letter we apply for the first time the `Hamiltonian formalism'
\cite{Henkel,Gunther} to the contact process with disorder.
This approach has the advantage that it is ideally suited to study
the asymptotic time regime, which due to the slow relaxation properties
of the model \cite{Webman} cannot easily be studied with simulations.
Since the contact process has the property of duality \cite{Liggett}, late and
early time behaviour can be related, provided suitable initial
conditions are chosen. In this way, the Hamiltonian approach can
also give insight in the spreading from a seed.

Consider the contact process on an open chain with $L$ sites.
The generator of the Markov process is given by the `Hamiltonian'
\begin{eqnarray}
H_{CP} = \sum_{k=1}^{L}
\mu_{k} M_{k}
+ \sum_{k=1}^{L-1} \frac{\lambda_{k}}{2} (n_{k} R_{k+1} + R_{k} n_{k+1})
\label{1}
\end{eqnarray}
in terms of the matrices
\begin{eqnarray*}
M = \left(\begin{array}{rr}
0 & -1\\
0 & 1\end{array}\right),
n = \left(\begin{array}{rr}
0 & 0\\
0 & 1\end{array}\right),
R = \left(\begin{array}{rr}
1 & 0\\
-1 & 0\end{array}\right)
\end{eqnarray*}

It is well known\cite{Gunther} that the steady state
probability distribution of a stochastic process coincides with the 
ground state of
its `Hamiltonian' and relaxation properties can be determined
from its low lying spectrum.
Although many of the
Hamiltonians associated with stochastic systems are non-hermitian,
techniques developed in the study of
quantum spin chains can often be successfully applied
to their stochastic counterparts\cite{Gunther,Dickper,Enrico,HV}.

Here we apply a strong disorder renormalization
group (RG) approach to the random contact process and the results are
compared with numerical estimates calculated using
exact and density matrix renormalization (DMRG) based diagonalisations.
We first announce our basic results which are summarized in the
schematic phase diagram in Fig. 1 as a function of the control
parameter
$A=[\ln \lambda ]_{\rm av}$ and the strength of disorder,
$R=[(\ln \lambda)^{2}]_{\rm av}-[\ln \lambda ]^{2}_{\rm av}$,
where, for the sake of simplicity we put $\mu_k=1$ (here
we use $[ \cdot ]_{\rm av}$ to denote averaging over quenched
disorder, while $\langle \cdot \rangle$ stands for
the average in the steady state). In the inactive phase,
for $A<A_c(R)$ the particle density
$\rho \equiv \lim_{L \to \infty} \frac{1}{L} \sum_{k}
[\langle n_{k} \rangle]_{\rm av}$ is zero, whereas in the active phase,
for $A>A_c(R)$, $\rho>0$. Close to the transition point, for
$\Delta=(A - A_c)/ A_c
\downarrow 0$, the particle density is singular,
$\rho \sim \Delta^{\beta}$ and the
correlation length $\xi$ and relaxation time $\tau$ diverge as $\xi
\sim |\Delta|^{-\nu_{\perp}},\ \tau \sim
|\Delta|^{-\nu_{\parallel}}$. The anisotropy exponent is
defined as $z=\nu_{\parallel}/\nu_{\perp}$. For a semi-infinite
system an exponent $\beta_{s}$, that describes the singular behavior of
the particle density at the surface, can be introduced \cite{denen}.

We found that, in one dimension, the universality class of the
phase transition depends on the strength of disorder, $R$, and
as indicated in Fig. 1 there are two different regions of critical
behavior. For $R>R_{c}$, the critical properties
of the system are controlled by a strong disorder fixed point,
which is situated at $R=\infty$, and in which the critical exponents are
conjecturedly exact:
\begin{equation}
\beta^{(\infty)}=\frac{3-\sqrt{5}}{2},\quad\beta_s^{(\infty)}=1,
\quad \nu_{\perp}^{(\infty)}=2.\;
\label{SDexp}
\end{equation}
This fixed point is characterised by strongly anisotropic scaling
\begin{eqnarray}
\ln \tau \sim \xi^{\psi}, \ \ \ \ \psi=\frac{1}{2} 
\label{2002}
\end{eqnarray}
Thus the anisotropy exponent, $z^{(\infty)}$ is formally infinity. 

For $R\leq R_{c}$, as shown
in Fig. 1, there is a line of
disorder dependent random fixed points (with finite $z$), in which 
the critical exponents vary
continuously with $R$. This behavior is consistent with an extension
of the Harris criterion for stochastic systems, which predicts a cross-over
exponent in the non-random system fixed point
as $y_{dis}(0)=2/\nu_{\perp}(0)-1>0$
\cite{MarDic,duitser}. Indeed, applying a real space RG
method \cite{HV} we have calculated $y_{dis}(0)$ in good agreement with the
Harris criterion.
\begin{figure}
\vskip 0.5 truecm
\centerline{
\psfig{file=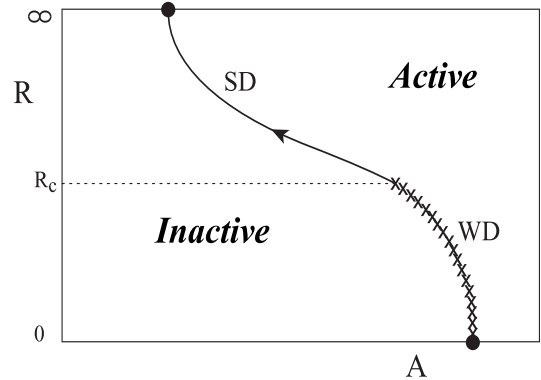, height=5.cm,width=7.cm}}
\vskip 0.2 truecm
\caption{Schematic phase diagram of the random contact process as a function
of $A$ and $R$
(see text). The inactive and active phases are separated by
a phase transition line, along which the critical exponents are $R$ dependent
for weaker disorder (WD, $ R \leq R_{c}$), or are determined by a 
strong disorder fixed point at $R=\infty$,
for strong disorder (SD, $R>R_{c}$).}
\end{figure}

In the strong disorder RG, which was introduced
\cite{MDH} for random antiferromagnetic spin
chains and which was later also applied to other one- and higher-dimensional
systems\cite{DF,SM,FerGrif}, one puts
the strength of the couplings and fields in the Hamiltonian (i.e. the
rates in the stochastic problem) in descending order. The strongest
one, denoted by $\Omega$, sets the energy scale in the system
and is decimated out and the neighboring rates are replaced by (generally)
weaker ones, obtained by a perturbation calculation.

In the random contact process, if the fastest rate is
$\lambda_{i}=\Omega$, then
the sites $i$ and $i+1$ will either be
both occupied or both empty almost always, and can thus be interpreted
as an effective two-site cell state. The effective decay
rate, $\mu'$, to go from $AA$ to $\emptyset\emptyset$ can then be
determined by applying perturbation theory to the Hamiltonian
(\ref{1}).
At the same time the effective
moment of the cell state, $n'$, is given by $n'=n_i+n_{i+1}$, where
in the starting situation $n_k=1, \forall k$.
Conversely, if the fastest rate
is $\mu_{i}=\Omega$, site $i$ will almost always be empty,
and in a perturbative treatment can be eliminated from the system.
This allows the determination of an effective coupling $\lambda'$ 
between the occupied site $i-1$ and
the empty site $i+1$.
In terms of $\tilde{\lambda}=\lambda/\sqrt{8}$ we obtain the
decimation equations ($\kappa=\sqrt{2}$):
\begin{eqnarray}
\mu' = \kappa \frac{\mu_{i} \mu_{i+1}}{\tilde{\lambda}_{i}},
\quad n'=n_i+n_{i+1},
\ \ \ \ \ \tilde{\lambda}' = \kappa
\frac{\tilde{\lambda}_{i-1}\tilde{\lambda}_{i}}{\mu_{i}}\;.
\label{4}
\end{eqnarray}

This decimation procedure is repeated and for strong
enough disorder, when the probability of generating a new rate
larger than $\Omega$ is negligible, the energy scale is continuously lowered
and at the same time the probability distributions,
$\pi(\mu,\Omega)$ and $P(\lambda,\Omega)$ approach their fixed point
forms. This RG procedure is easy to implement
numerically. However, in the 1d case the fixed point distributions
can be calculated analytically, using the results of a similar
analysis for 1d random quantum spin chains\cite{DF,SM,FerGrif}.
Here we mention that according
to Eq.(\ref{4}) the renormalised log-$\mu$ rates after large $L$ 
decimation steps are
expressed as the sum of $L+1$ original random log-$\mu$ rates
minus the sum of $L$ original random log-$\tilde{\lambda}$ rates 
(plus a negligible
constant proportional to $\ln \kappa$). Consequently
the fixed point, where the distributions $\pi(\mu,\Omega)$ and
$P(\tilde{\lambda},\Omega)$ are identical, is located at
$[\ln \mu]_{\rm av}=[\ln \tilde{\lambda}]_{\rm av}$.
Furthermore, at the
fixed point the
central limit theorem implies for the log-energy scale $-\ln \Omega 
\sim L^{1/2}$,
which is equivalent to (\ref{2002}).
As a consequence the probability distributions of the rates are
broadened without limit during renormalization. Thus the ratio of typical
neighboring rates goes to either zero or infinity. Therefore, in the
strong disorder fixed point the decimation equations in Eq.(\ref{4}) become
asymptotically exact. Thus the critical exponents 
in this fixed point (see (\ref{SDexp})),
which can be deduced from the analysis of the RG equations along
similar lines as for the random transverse Ising model \cite{DF},
are presumably exact, too.



In the following we check these predictions by numerical calculations
on lattices with $L$ up to $32$ with the DMRG method.
For the disorder we used a bimodal distribution with $\mu_k=1$ and
$P(\lambda)=[\delta(\lambda-\lambda_{+})+\delta(\lambda-\lambda_{-})]/2$,
where $\lambda_{\pm}=\exp(A\pm\sqrt{R})$, so that $A$
and $R$ are the parameters used in Fig. 1.
For smaller sizes, $L \le 14$, we performed exact
disorder averages, whereas for larger $L$-s we considered at least
10000 samples for each size.

\begin{figure}
\centerline{
\psfig{file=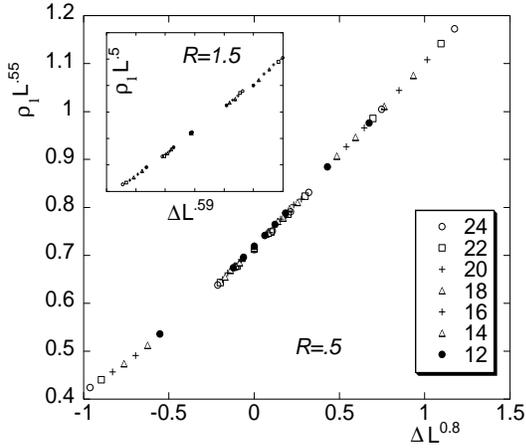, height=6.cm, width=7.cm}}
\vskip 0.2truecm
\caption{Scaling plot of the surface particle density
at $R=0.5$ and at $R=1.5$ (inset).}
\end{figure}

In the actual calculation we set $\mu_L=0$ and determine the average
particle density profile, $\rho_k=[\langle n_k \rangle]_{\rm av}$, so that
the order-parameter in the bulk (at the surface) is defined as
$\rho=\rho_{L/2}$ ($\rho_s=\rho_1$). Close to the critical point,
the densities obey scaling laws, such as
$\rho(L,\Delta)=L^{-x} \tilde{\rho}(\Delta L^{1/\nu_{\perp}})$, where
$x=\beta/ \nu_{\perp}$. A similar relation holds for $\rho_s$,
with the surface exponent $x_s=\beta_s/\nu_{\perp}$. From an optimal
data collapse, as illustrated in Fig. 2, the position of the transition
point, $A_c$, and the critical exponents are calculated.

For varying disorder strength, $R$, the estimated critical exponents,
$x$ and $x_s$, are plotted in Fig. 3, where one can observe the two
regions announced in Fig. 1. The crossover occurs near
$R_{c} \approx 1.5$. For higher disorder strenghts
the critical exponents seem to saturate at the
values given in (\ref{SDexp}). The estimate for $\nu_{\perp} = 1.7 \pm .3$ at $R=1.5$ is
also consistent with $\nu_{\perp}^{(\infty)}$. The error on this 
exponent is larger because of stronger finite size corrections.

\begin{figure}
\centerline{\psfig{file=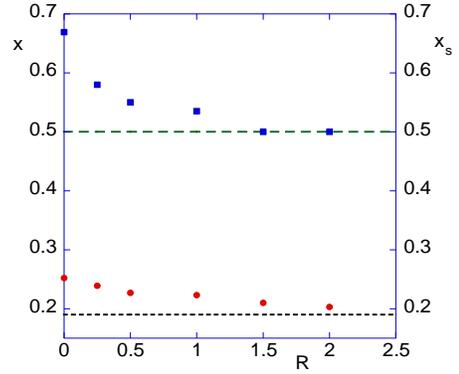, height=5.cm, width=6.cm}}
\vskip 0.2truecm
\caption{Numerical estimates of the exponents $x=\beta/\nu_{\perp}$
(circles)
and $x_{s}$ (squares). The broken lines indicate the
value at the strong disorder fixed point. The errors are of the same order as the
size of the symbols.}
\end{figure}

We note that the same type of weak-to-strong disorder cross-over scenario
has been observed in a class of random quantum spin chains\cite{EnricoFerenc}.
For these models, the strong disorder RG analysis leads to decimation equations of the
same form as Eq.(\ref{4}), with also $\kappa>1$. For these
$\kappa$-values, some of the
generated couplings (rates) can be larger than the decimated one. For weak
enough disorder, this happens so often that the RG approach
does not work. For stronger, but still finite disorder, however, such type of
decimation steps are rare so that during renormalization the system flows
into the strong disorder fixed point \cite{EnricoFerenc}.

We have also studied the dynamical scaling of the model by calculating the
probability distribution of the gap $\Gamma$ between the
groundstate and the first excited state of $H_{CP}$. In the small $\Gamma$
limit this distribution behaves as $P(\ln \Gamma) \sim \Gamma^{1/z'}$. Here
$z'$ is the disorder induced dynamical exponent, and the true dynamical
exponent of the system is given by $z={\rm max}(z',z(0))$. In the 
strong disorder fixed point
$z'$ is formally infinity and the scaling behavior of the gap distribution is
given by $P(\ln \Gamma) = \tilde{P}\left( \ln \Gamma/\sqrt{L}
\right)/\sqrt{L}$.
As shown in Fig. 4 the above scaling form is indeed satisfied in the 
strong disorder
region at $R=1.5$.

\begin{figure}
\centerline{\epsfig{file=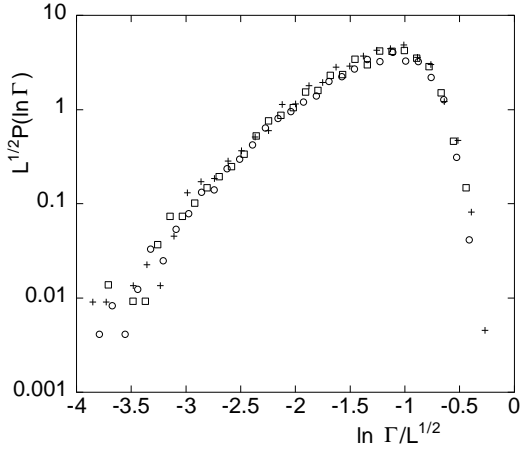, height=6.cm, width=7.cm}}
\vskip 0.2truecm
\caption{Scaling of the gap to the first excited state at
$R=1.5$ and $A=A_{c}=1.18$ ($L=16\ (+),20\ (\Box),
24\ \bigcirc$).}
\end{figure}

Based on the decimation rules in Eq.(\ref{4}), which
are very similar to that of the random transverse Ising model, we expect
that even in higher dimensions these random stochastic models
will renormalize into the strong disorder fixed point
for the higher dimensional random transverse Ising model \cite{2dRTIM}. Indeed from
\cite{2dRTIM} we obtain $\nu_{\perp}=1.07\pm.15$ and
$\beta=1.07\pm.12$, consistent with the values obtained for
a contact process with sufficiently strong dilution \cite{weerDick}. For smaller
values of the dilution, $\beta$ seems to vary continuously
\cite{weerDick}, indicating
that a scenario similar to that in $d=1$ is valid.
Moreover, the breakdown of dynamical scaling observed in
\cite{weerDick} can be related to the strongly anisotropic
scaling (\ref{2002}). Using $\psi=.42\pm.06$ \cite{2dRTIM}, it is
even possible to predict \cite{JFC2} the values of some of the dynamic
exponents introduced in \cite{weerDick}. 

In conclusion, we have found that the critical properties of absorbing states
models in the directed percolation universality class in the presence of random transition
rates are controlled by a strong disorder fixed point, if the disorder is sufficiently strong.
The exponents associated with this transition,
are conjectured exact. This is the
first example of an absorbing state phase transition with quenched disorder
for which exact
information becomes available.
A more extensive paper on our results will be published elsewhere
\cite{JFC2}.


F.I. is grateful to P. Grassberger, J.D. Noh and F. van Wijland for
stimulating discussions and to L. Gr\'an\'asy for his help in the numerical
calculations. F.I.'s  work has been supported by the Hungarian National
Research Fund under  grant No OTKA TO34183, TO37323,
MO28418 and M36803, by the Ministry of Education under grant No FKFP 87/2001,
by the EC Centre of Excellence (No. ICA1-CT-2000-70029) and the numerical
calculations by NIIF 1030. J.H. is a research `aspirant' of
the FWO-Vlaanderen.
\vskip -.4cm

\end{document}